\documentclass{revtex4}

\usepackage{epsfig}

\begin{document}

\title{On the contact conditions for the charge profile in the theory of the
electrical double layer for nonsymmetrical electrolytes.}
\author{M. Holovko}
\surname{Holovko}
\affiliation{Institute for Condensed Matter Physics,
National Academy of Sciences\\1 Svientsitskii Str., 79011 Lviv, Ukraine\\}
\author{D. di Caprio}
\surname{di Caprio}
\affiliation{Laboratoire d'\'{E}lectrochimie et Chimie Analytique, UMR 7575,\\
UPMC Univ Paris 6, CNRS, ENSCP\\
Universit\'e P. et M. Curie, B\^at. F74, B.P. 39\\
4, Place Jussieu, \ 75230 Paris Cedex 05, France\\}

\begin{abstract}
The contact value of the charge profile for nonsymmetrical electrolytes
is presented as the sum of three contributions. One of them is the normal
component of the Maxwell electrostatic stress tensor.
The second one is the surface electrostatic property defined as the integral of
the product of the gradient of the electrical potential and the density
distribution function of coions.
The third term is the bulk contribution defined by the sum for anions and for
cations of the product of their charge and their partial pressure.
For noncharged surfaces only the last two are present and have the same sign in
the case of size asymmetry. In the case of charge asymmetry the contact value of
the charge profile is the result of the competitions of bulk and surface terms
in which the bulk term is dominant. Using both the contact theorems for the
density and the charge profiles, the exact expressions for the contact values of
the profiles of coions and counterions are obtained and some related properties
are discussed.

\end{abstract}

\maketitle
The properties of the electrical double layer are described by the density
and charge profiles $\rho(z)$ and $q(z)$. They are connected with the density
distribution functions $\rho_{\alpha}(z)$ for ions of type $\alpha$ having
the charge $e_{\alpha}$ by the relations
\begin{eqnarray}
  \rho(z) &=& \sum_{\alpha}\rho_{\alpha}(z)\\
     q(z) &=& \sum_{\alpha}e_{\alpha}\rho_{\alpha}(z)
\end{eqnarray}
where $z$ is the normal distance from the wall.

For the contact value (CV) of the density and charge profiles, for the primitive
model of electrolyte solutions, exact relations have established known as the
contact theorems (CTs).
The CT for the density profile states that the CV of the
number density profile $\rho^{ct}$ is given by the sum of the
bulk electrolyte pressure $P$ and the electrostatic Maxwell stress tensor contribution
\cite{HendersonBlum,HendersonBlumLebowitz}
\begin{eqnarray}\label{eq:CT1}
  \rho^{ct} \equiv \rho(d/2)= \beta P + \beta \frac{\varepsilon E^2}{8\pi}
\end{eqnarray}
where $d$ is the ion diameter, assumed to be the same for all species,
$\beta =1/(k_B T)$ with $k_B$ the Boltzmann constant and $T$ the absolute
temperature, $\varepsilon$ is the dielectric constant of the solvent,
$\varepsilon E / 4\pi = q_s$ is the surface charge density per unit area on the
wall. The CT for the charge profile was formulated recently by
Holovko et al. \cite{MHJPBDDC}. In contrast to the theorem for the density profile,
the CV of the charge profile $q^{ct}$ has a nonlocal character and in
general can be presented in the form
\begin{eqnarray}\label{eq:CT2}
  q^{ct} \equiv q(d/2)=\beta \int_{d/2}^\infty \hat{\rho}(z) \frac{\partial \psi (z)}{\partial z}
   dz + \beta \sum_{\alpha} e_{\alpha} P_{\alpha}
\end{eqnarray}
where $\hat{\rho}(z) = \sum_{\alpha} e^2_{\alpha} {\rho}_{\alpha}(z)$,
$\psi(z)$ is the electrical potential
\begin{eqnarray}
  \psi(z) = -\frac{4\pi}{\varepsilon} \int_z^\infty q(z_1) (z_1-z) dz_1
\end{eqnarray}
and $P_{\alpha}$ is the bulk partial pressure for ions of type $\alpha$.

Both CTs (\ref{eq:CT1}) and (\ref{eq:CT2}) can be obtained by
direct integration of the system of Born-Green-Yvon (BGY) equations for the
singlet distribution functions $\rho_{\alpha}(z)$
\cite{MHJPBDDC,CarnieChan}.
This procedure can be easily repeated for an electrolyte with different ion
diameters $d_{\alpha}$ and leads to changing only $\rho^{ct}$ and $q^{ct}$
in relations (\ref{eq:CT1}) and (\ref{eq:CT2}) by the corresponding
CVs given by
\begin{eqnarray}
  \rho^{ct} &=& \sum_{\alpha} \rho_{\alpha}(d_{\alpha}/2)\\
  q^{ct} &=& \sum_{\alpha} e_{\alpha} \rho_{\alpha}(d_{\alpha}/2)
\end{eqnarray}
and the bulk partial pressure $P_{\alpha}$ according to the virial expression
for the bulk pressure of a ion hard sphere system \cite{MarchTosi}
is given by
\begin{eqnarray}
  \beta P_{\alpha} =\rho_{\alpha}\left(
    1+\frac{2}{3}\pi \sum_{\gamma}\rho_{\gamma} d_{\alpha\gamma}^3
    g_{\alpha\gamma}(d_{\alpha\gamma}^+)\right) +\frac{1}{3}\beta E_{\alpha}^{Coul}
\end{eqnarray}
where $d_{\alpha\gamma}= (d_{\alpha} + d_{\gamma})/2$,
$g_{\alpha\gamma}(d_{\alpha\gamma}^+)$ is the CV of the pair
distribution function $g_{\alpha\gamma}(r)$ and
\begin{eqnarray}
  E_{\alpha}^{Coul} = \frac{1}{2} \rho_{\alpha} \sum_{\gamma} 4\pi \rho_{\gamma}
    \frac{e_{\alpha}e_{\gamma}}{\varepsilon} \int_0^\infty g_{\alpha\gamma}(r)
    r dr \label{eq:Ecoul}
\end{eqnarray}
is the coulombic part of the partial internal energy.

For the symmetrical electrolyte $\hat{\rho}(z) = e^2 \rho(z)$ ($e$ being
the elementary charge), it was shown \cite{MHJPBDDC2} that the CV
of the charge profile can be presented as the sum of the normal component of the
Maxwell electrostatic stress tensor and a new electrostatic property defined by the
integral of the product of the gradient of the electrical potential and the singlet
distribution function of coions (ions having the identical sign as the surface
charge).
In this note we generalize this result for nonsymmetrical electrolytes and
discuss some specific properties connected with the charge and size asymmetry of
cations and anions.\\

First, we consider a nonsymmetrical electrolyte with a single type of anion
and of cation, $\rho_\pm$, $e_\pm$ and
$d_\pm$ are respectively the densities, charges and diameters of the ions and
$\hat{\rho}(z)$ can be written as
\begin{eqnarray}
  \hat{\rho}(z) &=& e^2_+ \rho_+(z) + e^2_- \rho_-(z)\\
                &=&e_\mp q(z)-e_\pm\rho_\pm(z)(e_\mp-e_\pm)\label{eq:hatrhorep}
\end{eqnarray}
where the upper/lower sign are relative to a positively/negatively charged surface.
As a result the CT for the charge profile for nonsymmetrical electrolyte
can be presented in the following form
\begin{eqnarray}
         q^{ct}  &=& \beta e_\mp \frac{\varepsilon E^2}{8\pi} - 
      \beta e_\pm(e_\mp - e_\pm) \mathcal{J}_\pm + \beta (e_+P_++e_-P_-) \label{eq:CTI1}
\end{eqnarray}
where 
\begin{eqnarray}
  \mathcal{J}_\pm = \int_{d_\pm/2}^\infty dz \frac{\partial \psi (z)}{\partial z} \rho_\pm(z)
   \label{eq:Idef}
\end{eqnarray}
$\rho_\pm(z)$ is the density distribution function for coions near a
positively/negatively charged surface.
The result obtained is the generalization of our previous result \cite{MHJPBDDC2}
for the CV of the charge profile of the symmetrical electrolyte to the
nonsymmetrical case. Similar as for the symmetrical electrolyte expression 
(\ref{eq:CTI1}) has a simple interpretation. The difference between forces from
counterions and coions required to obtain $q^{ct}$ is connected with three
contributions: the electrostatic Maxwell stress tensor contribution, the
electrostatic contribution from coions described by the electrostatic property
$\mathcal{J}_+$ or  $\mathcal{J}_-$ depending on the sign of the surface charge
and a special combination of the partial pressures for coions and counterions.

From the CTs for the density and charge profiles presented in their
form (\ref{eq:CT1}) and (\ref{eq:CTI1}), it is possible to obtain
the following exact relations for the CVs of the density profiles
for coions and counterions respectively
\begin{eqnarray}
  \rho_{+}^{ct}\equiv\rho_+(d_+/2) &=& \beta P_+ + \beta e_+ \mathcal{J}_+,  \nonumber\\
  \rho_{-}^{ct}\equiv\rho_-(d_-/2) &=&
    \beta P_- + \beta \frac{\varepsilon E^2}{8\pi} - \beta e_+ \mathcal{J}_+ 
\end{eqnarray}
for a positively charged surface or
\begin{eqnarray}
  \rho_{+}^{ct}\equiv\rho_+(d_+/2) &=&
    \beta P_+ + \beta \frac{\varepsilon E^2}{8\pi} -  \beta e_- \mathcal{J}_-,  \nonumber\\
  \rho_{-}^{ct}\equiv\rho_-(d_-/2) &=& \beta P_- + \beta e_- \mathcal{J}_-
\end{eqnarray}
for a negatively charged surface.
Since the CV of the density profile of coions cannot be negative,
we also have the following conditions
\begin{eqnarray}
  P_+ +e_+ \mathcal{J}_+ \ge 0
\end{eqnarray}
for a positively charged surface or
\begin{eqnarray}
  P_- +e_- \mathcal{J}_- \ge 0
\end{eqnarray}
for a negatively charged surface.

These inequalities reduce to equalities in the limit $E \rightarrow \infty$
and as a consequence we have an exact cancellation of $e_+\mathcal{J}_+$
(or $e_-\mathcal{J}_-$) by the partial pressure $P_+$ (or $P_-$).
Similar as for the symmetrical electrolyte in the limit of small $E$, $\mathcal{J}_+$ (or $\mathcal{J}_-$)
has a linear dependence on $E$ and we can conjecture that $\mathcal{J}_+$ (or $\mathcal{J}_-$)
is a monotonic function of $E$.\\

However, for a noncharged surface ($E=0$) and nonsymmetrical electrolytes,
the charge profile $q(z)$ is not zero in contrast to the symmetrical case.
The charge and/or size asymmetry of cations and anions induces a double layer
with an associated surface potential \cite{TorrieValleau,Valiskoetal}.
According to the CT eq.~(\ref{eq:CTI1}), the CV of the
charge profile for a noncharged surface is defined by two terms
\begin{eqnarray}
  q^{ct}
            &=& -\beta e_+(e_--e_+) \mathcal{J}_+ + \beta (e_+P_+ + e_-P_-)\label{eq:CTIE0a}\\
            &=& -\beta e_-(e_+-e_-) \mathcal{J}_- + \beta (e_+P_+ + e_-P_-).\label{eq:CTIE0b}
\end{eqnarray}
Both forms for the CV of the charge are equivalent and due to this for a
noncharged surface we have the following symmetrical relation
\begin{eqnarray}
  e_+\mathcal{J}_+ + e_- \mathcal{J}_- = 0.
\end{eqnarray}

We may now distinguish the electrical double layer due to the charge and size
asymmetry of ions for they are different in nature.\\
The influence of size asymmetry has a short-ranged steric origin since the
smallest ion can approach the surface more closely than the larger ions.
If we consider for simplicity a charge symmetrical electrolyte ($e_+=-e_-=e$)
according to relation (\ref{eq:CTIE0a}) 
\begin{eqnarray}
  q^{ct} &=& 2\beta e^2 \mathcal{J}_+ + \beta e (P_+ - P_-). \label{eq:CTIE02}
\end{eqnarray}
Due to electroneutrality, the first layer constituted of the smaller ions has
to be compensated by the following layer where we find the larger ions. Then
intuitively, the CV of the profile for the larger ions which are further
from the wall has to be larger in order to compensate the profile for the
smaller ions which starts before as they can come closer to the wall.
For example, for the case $d_+<d_-$ we will have $\rho_{+}^{ct} < \rho_{-}^{ct}$
and $q^{ct}$ will be negative.
This intuitive result is indeed in agreement with the CT for the
charge given in eq.~(\ref{eq:CTIE02}).
For this case $(d \psi(z)/d z)<0$ and according to definition
eq.~(\ref{eq:Idef}) $\mathcal{J}_+<0$ thus the first term of
eq.~(\ref{eq:CTIE02}) is negative. The second term at least for small
ion concentration is also negative since the dominant contribution
to the partial pressure comes from the Coulombic part of the partial
internal energy eq.~(\ref{eq:Ecoul}) which is more negative for smaller ions.
This means that in this case $q^{ct}$ will be negative in agreement with the
intuitive statement.
For the opposite case $d_+>d_-$, $\rho_{+}^{ct} > \rho_{-}^{ct}$
and $q^{ct}$ will be positive. This result
also corresponds to the CT~(\ref{eq:CTIE02}) as
in this case $(d \psi(z)/d z)>0$, $\mathcal{J}_+>0$ and in parallel $P_+>P_-$
and $q^{ct}$ will be positive.\\
The influence of the charge asymmetry on $q_{ct}$ is different. It is associated
with the electrostastics and results from the long-range behaviour of the
profiles of the cations and anions leading to an exclusion effect by the neutral
surface \cite{TorrieValleau,Valiskoetal}.
From \cite{DDCJSJPB}, when the conditions are such that electrostatic
interactions become important in comparison to steric effects, we know for
symmetrical electrolytes that the density profile decreases when we approach the
wall producing a negative adsorption of ions.
This effect is more pronounced the higher the charge of the ions.
Now this phenomenon can be generalized to ions with different charges where to
simplify we consider size symmetrical electrolytes ($d_+=d_-=d$).
And as a consequence, the CV of the charge for the ion carrying the
higher charge will be smaller
resulting in a CV of the charge profile negative for highly
charged cations and positive for highly charged anions.
For example in the case $e_+> |e_-|$, $e_+\rho_+^{ct}< |e_-|\rho_-^{ct}$ and
$q^{ct} \le 0$.
In this case $(d \psi(z)/d z)>0$, $\mathcal{J}_+>0$
and the first term of eq.~(\ref{eq:CTIE0a}) is positive.
The second term in expression (\ref{eq:CTIE0a}) is negative
since for highly charged ions the coulombic part of the partial internal
energy is more negative than for weakly charged ions and thus
$e_+P_+ + e_-P_-<0$.
However, this term must be sufficiently negative in order to partially cancel
the first positive term in the expression of $q^{ct}$ and obtain an overall
negative value for $q^{ct}$.
For the case $e_+< |e_-|$, $q^{ct}$ will be positive. In this case,
$(d \psi(z)/d z)<0$, $\mathcal{J}_+<0$ and the first term of $q^{ct}$ is
negative.
The second term is positive and again it must cancel partially the first negative
term in order to have $q^{ct}\ge 0$.
We can see that the sign of $q^{ct}$ for the charge asymmetrical case is
the result of the partial compensation of the bulk term, which is the sum of the
product of ionic charges with the partial pressure of ions, and the surface
term connected with the electrostatic property $\mathcal{J}_+$.
Both contributions have opposite signs but the bulk contribution
must be dominant.\\

The simultaneous consideration of charge and size asymmetry effects is more
complex. Nevertheless, expressions (\ref{eq:CTIE0a}-\ref{eq:CTIE0b}) for the CV
of the charge profile are correct for this case also.
It is clear that in some situations these two effects can enhance each others' influence
and in other situations they can weaken or even cancel each other.\\

In this paper, results previously obtained for symmetrical
electrolytes \cite{MHJPBDDC2} are generalized for charge and/or size asymmetric
systems.  It is shown that the CV of the charge profile can be
presented as the sum of three contributions.
One of them is the normal component of the electrostatic Maxwell stress tensor.
The second one is the electrostatic contribution from coions, described by the
electrostatic property $\mathcal{J}_+$ or $\mathcal{J}_-$ depending on the sign
of the surface.
Similar as for the symmetrical case \cite{MHJPBDDC2}, $\mathcal{J}_+$ or
$\mathcal{J}_-$ are defined by the integral from the product of the gradient of
the electrical potential and the density distribution function of coions near
the charged surface.
The third term is the bulk contribution defined by the sum of the product of the
charges of the coions and counterions multiplied by their corresponding partial
pressures.
Using the CTs for the charge and density profiles, the exact
expressions for the CVs of the density profiles for coions and
counterions are presented and discussed.
We have also discussed the influence of the charge and size asymmetry on the CV
of the charge profile for a noncharged surface.
For noncharged surfaces, from the CT for the charge, the CV of the
charge profile now only has two contributions, the Maxwell tensor
being zero.
We show that in the case of size asymmetry both contributions have the same sign
which corresponds to the sign of the CV of the charge profile.
In the case of charge asymmetry the two contributions have opposite signs.
However the bulk contribution is dominant and imposes its sign to the CV
of the charge.\\

Finally, the generalization of the results we have obtained for the more than
two components case faces the principal difficulty related with the problem of
expressing $\hat{\rho}(z)$ using the charge profile $q(z)$ and the coion density
functions as in eq.~(\ref{eq:hatrhorep}).  This problem can be solved only for
the electrolyte with a single type of counterions.  In this case, for example,
for the positively charged surface
\begin{eqnarray}
  q^{ct} &\equiv& \sum_{\alpha} e_{+\alpha} \rho_{+\alpha}(d_{+\alpha}/2) + e_- \rho_-(d_-/2)\nonumber\\
   &=&
    \beta e_- \frac{\varepsilon E^2}{8\pi}
  - \beta \sum_{\alpha} e_{+\alpha} (e_--e_{+\alpha}) \mathcal{J}_{+\alpha} + \beta e_- P_-
  + \sum_{\alpha} e_{+\alpha} P_{+\alpha} \label{eq:CT4}
\end{eqnarray}
where 
\begin{eqnarray}
  \mathcal{J}_{+\alpha} = \int_{d_{+\alpha}/2}^\infty \rho_{+\alpha}(z) \frac{\partial \psi (z)}{\partial z} dz.
\end{eqnarray}
From eq.~(\ref{eq:CT1}) and (\ref{eq:CT4}) we can obtain that
\begin{eqnarray}
  \sum_{\alpha}(e_--e_{+\alpha})\rho_{+\alpha}(d_+/2) =
      \beta \sum_{\alpha}(e_--e_{+\alpha}) P_{+\alpha} +
      \beta \sum_{\alpha}(e_--e_{+\alpha}) e_{+\alpha} \mathcal{J}_{+\alpha}.
\end{eqnarray}
However this equation is not sufficient to obtain the expressions for the CV of
the ionic density profiles in a similar way as starting from
expression~(\ref{eq:CTI1}).

\begin{acknowledgments}
This work has been carried out in the framework of the agreement
n\symbol{23}18964 between the French Centre National de la Recherche
Scientifique (CNRS) and the National Academy of Sciences of Ukraine (NASU).
The authors express their gratitude to J.P. Badiali for the support of this work
and very useful discussions.
\end{acknowledgments}

\renewcommand{\baselinestretch}{1.55} \small \normalsize
\bibliographystyle{apsrev}
\bibliography{paper}

\end{document}